# New care pathways for supporting transitional care from hospitals to home using AI and personalized digital assistance

Ionut Anghel[1*], Tudor Cioara[1*], Roberta Bevilacqua[2], Federico Barbarossa[2], Terje Grimstad[3], Riitta Hellman[3], Arnor Solberg[4], Lars Thomas Boye[4], Ovidiu Anchidin[5], Ancuta Nemes[5] and Camilla Gabrielsen[6]

[1] Computer Science Department, Technical University of Cluj-Napoca, G. Barițiu 26-28, 400027 Cluj-Napoca, Romania,

[2] Istituto Nazionale di Ricovero e Cura per Anziani, Via S. Margherita 5, 60124 Ancona, Italy

[3] Karde AS, Irisveien 14, 0870 Oslo, Norway

[4] Tellu AS, Lysaker Torg 15, 1366 Lysaker, Norway

[5] Niculae Stancioiu Heart Institute, Moților 19-21, 400001 Cluj-Napoca, Romania

[6] Farsund Kommune, Postboks 100, 4552 Farsund, Norway

*Corresponding Authors: ionut.anghel@cs.utcluj.ro; tudor.cioara@cs.utcluj.ro

**Abstract.** Transitional care may play a vital role for the sustainability of Europe's future healthcare system, offering solutions for relocating patient care from hospital to home therefore addressing the growing demand for medical care as the population is ageing. However, to be effective, it is essential to integrate innovative Information and Communications Technology (ICT) technologies to ensure that patients with comorbidities experience a smooth and coordinated transition from hospitals or care centers to home, thereby reducing the risk of rehospitalization. In this paper, we present an overview of the integration of IoT, artificial intelligence, and digital assistance technologies with traditional care pathways to address the challenges and needs of healthcare systems in Europe. We identify the current gaps in transitional care and define the technology mapping to enhance the care pathways, aiming to improve patient outcomes, safety, and quality of life avoiding hospital readmissions. Finally, we define the trial setup and evaluation methodology needed to provide clinical evidence that supports the positive impact of technology integration on patient care and discuss the potential effects on the healthcare system.

## Introduction

As people get older, they may face a variety of health challenges and multimorbidity patterns that require hospitalization. While it's natural for seniors to want to return home as soon as possible, adjusting to life and following the prescribed treatment plan after hospitalization can be difficult. This makes the risk of readmission during their convalescence periods after discharge high. For example, in the case of heart failure, 25% of the patients require readmission within one month of hospital discharge, and 40% within 3 months after discharge [1]. Similarly, the 30-day rehospitalization rate is approximately 25% for patients with dementia, and around 17% for patients with diabetes [2]. The causes of this phenomenon are multifaceted, involving factors related to both patients and the healthcare system. The older adults' chronic condition is an important factor that contributes to their fragility, compounded by additional risk factors such as difficulty performing everyday tasks, social isolation and lack of support, cognitive issues when they become ill, and emotional anxiety. When patients receive care from various health care providers, in the municipalities, at facilities or at home without a unified care plan, a fragmented approach to transitional care is likely to occur. This fails to support patients in stabilizing and avoiding acute deterioration from chronic diseases, resulting in economic consequences due to high rates of preventable hospitalizations and emergency / ambulatory department visits.

Technology-assisted transitional care models can help the healthcare providers to better monitor patients' conditions, provide timely follow-up care, and communicate more effectively with patients [3]. They can help reduce the likelihood that a chronically ill older adult will need to be readmitted to the hospital, ultimately improving their health outcomes, and reducing healthcare costs. By using Internet of Things (IoT) sensors, Artificial Intelligence (AI) and Machine Learning (ML), it is possible to assess elderly patients before their discharge, stratifying the risks for hospital readmission [4]. The data analytics take advantage of infrastructures for remote monitoring of daily life activities, vital signs and medication intake to identify potential situations that may lead to rehospitalization. At the same time digital assistants can provide personalized recommendations post-discharge. However, to fully maximize technology potential, the redesign of the care pathways to integrate such technologies, establish data sharing requirements, and providing clinical evidence demonstrating the impact on reducing the risk of readmissions needs to be tackled.

In transitional care, the adoption of digital solutions is rather limited [5][6]. The Information and Communications Technology (ICT)-based solutions are fragmented and address only specific cases of the transitional process and lack coordination activities to bridge the transition from hospital to home [7]. Studies have shown that readmission can be prevented [8] and that for every four patients enrolled with ICT coaching and communication tools, one hospitalization is avoided while significant cost savings are obtained. Nevertheless, to achieve better outcomes and reduce healthcare utilization, there is a strong need for improved communication, data sharing, and trust among healthcare providers and patients [9]. Lack of holistic, multi-criteria assessment (health, social, economic, etc.) at discharge, poor communication with patients, inadequate planning, and community support can result in adverse outcomes including unnecessary hospital readmissions [10]. The patient follow-up after discharge often requires frequent home visits during convalescence, that can be costly and time consuming for health professionals [11]. The recent development of IoT sensors and wearable devices have the potential of improving this, by enabling the remotely monitor patients' health status in their homes [12]. Sensors can be used to track the adherence to medication plan, daily activities, or lifestyle changes, while communication platforms may facilitate proactive intervention and patient education at home [13]. However significant challenges related to adoption of transitional care should be addressed such as the integration and re-configuration of patient care pathways and workflows, lack of field support or inadequate healthcare providers training [14]. Finally, it is increasingly recognized the importance of community-based services in supporting patients' recovery and preventing readmissions [15]. But open issues around reimbursement practices, roles and reasonability sharing, inter-provider communication should be addressed.

In this context we can identify several important challenges and necessities for transitional care. *Effective communication, information sharing, and patient's remote follow-up* is the backbone for enacting transitional care processes. In this sense as presented above IoT, ML and digital assistance can lay the groundwork for objective monitoring of patient's healthcare data in their homes and effective distribution and sharing for all healthcare stakeholders (emergency departments doctors, formal and informal care givers, patients themselves), assuring the continuity of care. There is a strong need to *improve the care coordination by redesigning the care pathways to integrate modern ICT techniques*. This can be achieved by defining clear roles and responsibilities in effectively using the technology and the new digital channels to avoid fragmentation and lack of care coordination. *Financial sustainability of transitional care programs and associated reimbursement mechanisms* is challenging the implementation of such programs. New payment models need to be considered examining issues such as who is responsible for paying for healthcare services (national or local government, insurance companies, patients, etc.), and how hospitals/care centres are incentivized to invest in avoiding readmissions, given that they currently receive funding based on the number of admissions. Moreover, *limited availability and allocation of resources for the healthcare process in general and transitional care in special*. In Europe, healthcare systems face significant pressure due to limited resources, increase in cost of care, and significant burden of the growing burden of chronic diseases. To effectively manage the shortages in healthcare professionals and budgetary constraints, innovative tools and technologies are needed to enable effective remote follow-up of patients post discharge to reduce their rehospitalization rates.

In our view the solution for advancing transitional care is to research, develop and validate innovative IoT, ML, and digital assistance-based integrated solutions for enabling the continuity of care for patients in their transition from hospital to home by assessing patients before discharge for hospital readmissions risks, remote monitoring of daily life activities, vital signs, and medication to identify situations leading to rehospitalization and provide post-discharge personalized recommendations using digital assistants. This paper addresses the above open challenges and necessities by proposing a methodology and augmented care pathway for developing, validating and evaluating a technology-assisted transitional care solution based on IoT monitoring, ML, and digital assistance while considering the specific contexts of different healthcare systems in Europe as well as various types of comorbidities. The methodology will be designed and developed in the TransCare EU project [16] that promotes the integration of modern technologies and successful practices into diverse healthcare systems in Europe (Italy, Norway and Romania), validating their effectiveness in different contexts and with various comorbidities by setting up trials to provide solid clinical evidence that supports the adoption of transitional care solutions and their potential to reduce the need for hospital readmissions. The methodology seeks to improve transitional care: (i) effectiveness by providing the IoT infrastructure for remote monitoring of patient's data and adherence to treatment, (ii) efficiency by offering digital assistance based virtual communication and coordination for proactive and personalized intervention freeing up healthcare providers time (iii), timely care due to AI-driven proactive detection of lack adherence post-discharge leading to better outcomes, and more efficient care pathways and (iv) safety and equity of the process.

The main research questions addressed in our paper are:

- How can the transitional care process be personalized?
- What are the innovative hardware and software stack of resources needed for remote monitoring?
- How can the patient assessment and follow-up care processes be supported by technology?
- Can the post-discharge medication plan be managed using ICT technologies?
- How can family members be engaged through support and communication channels?
- Can such closed loop information system assure the continuity of care and avoid/delay rehospitalization?

The remainder of this paper is structured as follows: Methods Section presents the proposed methodology for transitional care innovation while Discussion Section analyses the potential impact in the healthcare domain and potential areas for future research.

## Methods

The first step before designing innovative technologies for transitional care is to study how the care pathways need to be re-designed for allowing the integration and usage of such technology in different hospitals and healthcare settings providing the necessary knowledge for care relocation form hospital to home. The boundaries of the transitional care process need to be set up by identifying medical conditions or situations that can be safely managed at home with the assistance of the proposed technology. This can lead to defining guidelines for shaping the responsibilities of the healthcare providers in the process, the associated technology supported communication channels and coordination procedures for professionals involved in the patient's care. Additionally, the value-based payment models in different countries need to be analysed to provide guidelines for shared payment responsibilities (national or local government, insurance companies, patients), and quality metrics focused on patient outcomes that can incentivize hospitals and healthcare providers to invest in avoiding readmissions through innovative technologies.

**Research hypothesis**

The proposed research process is focused on the following hypotheses:

**H1.** The adoption and usage of an IoT, ML, and digital assistance-based solution for transitional care can reduce the patients' rehospitalization rate. The ML models can analyse the patients' daily life and health data acquired using IoT devices, to identify patterns that precede hospital readmissions, proactive interventions and self-treatment planning can be delivered to prevent avoidable readmissions and improve transitional care outcomes.

**H2.** The integration and usage in transitional care of digital assistance-based virtual communication and remote monitoring and ML assessment of patient state can improve the healthcare system efficiency by reducing the follow-up time for healthcare providers. Healthcare providers can prioritize their resources and focus on patients who require immediate attention or intervention, receiving follow-up care more efficiently. Reduction of the stress burden on healthcare workers can be assessed with specific questionnaires such as Zarit Burden Interview [17].

**H3.** The patient remote monitoring, follow-up and digital assistance-based support can help them during their recovery process, address concerns promptly, asses their adherence to the treatment and make necessary adjustments to their care plans, thereby enhancing patient satisfaction and Quality of Life. Thus, we expect that the Quality of Life and overall satisfaction of patients assessed using validated tools such as the SF-36 [18] to be improved.

**H4.** The solution, through its holistic features, will reduce the cost of transitional care by identifying risks and preventing costly rehospitalizations, setting up early interventions that can prevent the progression of illnesses and reduce the need for expensive treatments, and ultimately lower healthcare costs. It can reduce the utilization of healthcare resources by eliminating redundant procedures, minimizing medication errors, and decreasing administrative burdens.

The hypothesis can be assessed in specific trials set ups in different countries, using a before and after strategy, assessing the baseline value before the technology introduction and after usage in trials.

**Enhancing the traditional care pathways**

The traditional care pathway models are focused on committing dedicated healthcare resources for assuring the continuity of care [19]. The existing evidence-based transitional care model promotes nurse-led intervention targeting older adults at risk while the patients and family caregivers' engagement is rather limited. This is an inefficient and unsustainable approach leading to waste of resources and poor quality and effectiveness of care. At the same time the transitional care process experiences many gaps in many places so there is an evident need to systematize the effectiveness of these processes. Recent advancements in IoT and sensors could revolutionize this scenario by allowing for the remote and daily tracking of crucial transitional care elements, such as adherence to prescribed therapies and lifestyle modifications, thus facilitating more coordinated interventions. Additionally, this technology has the potential to enhance patient engagement and adherence, as well as improve timely communication among all involved parties. Given their busy schedules and numerous responsibilities, doctors and caregivers are no longer interested in general-purpose care management information; instead, they seek personalized, near real-time, targeted data about their patients and loved ones.

Also, the there is a need for innovative services integrating solutions supporting the non-face to face post-discharge follow up of patients by doctors such as advanced communication tools with patients and caregivers, remote assessment of patient adherence to treatment and patient and/or family/caretaker support for self-management and independent living. Figure 1 shows a view of an enhanced care pathway model that brings a perspective shift proposing digital solutions and modern technologies integration in the care chain for seamless coordination of care to proactively detect problems thus avoiding re-hospitalization.

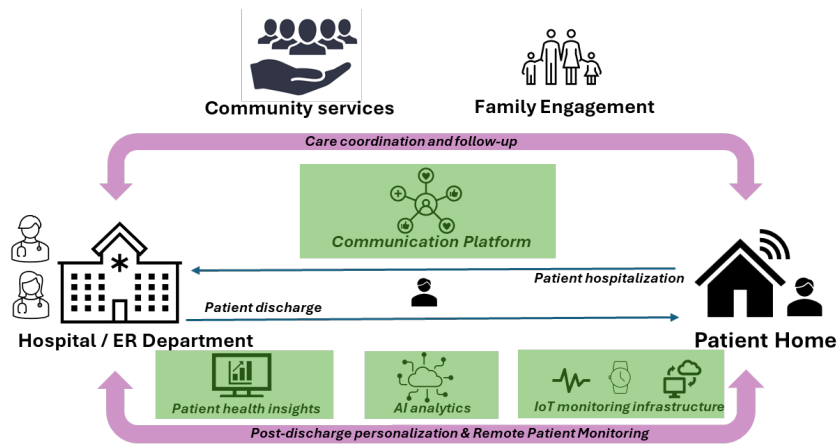

**Figure 1.** Care pathway enhancement.

Transitional care is differently implemented in European countries thus this raises the complexity of applying general models at European level.

Transitional care in **Italy** represents a significant challenge and opportunity for the national health care system. One emerging model is that of the "smart hospital," inspired by European examples, which is based on the idea of a hospital without walls, capable of reaching patients' homes directly through ICT. The main mission is to integrate hospital services with territorial services, creating a continuum of care that supports patients even outside traditional healthcare facilities. In Italy, there is a strong demand for territorial health services, but the response to this demand is often insufficient due to fragmentation and lack of resources at the municipal level. Although hospitals are mainly focused on acute care management, the transformation to smart hospital is essential to evolve toward a territorial healthcare system. To adequately meet this demand, territorial services (such as local health authorities) need to be strengthened and the specificity and expertise of staff caring for patients at home needs to be improved. Currently, these facilities often lack specialized staff by disease type, which reduces the effectiveness of home care. Another crucial aspect is investment in technologies that enable the hospital to enter patients' homes. The combination of social and health funds is essential to support this development. Social technologies, although less funded than health technologies, play an important role and must be integrated into the community care system. This includes the adoption of digital platforms and devices that, while not recognized as medical tools, contribute significantly to patient well-being. The issue of funding for these nonmedical devices, such as social robots or smart watches, is indeed relevant. Although they do not fall into the categories of medical devices, they fall into the sphere of social welfare and require a review of funding dynamics. In addition, intensive work on verifying the effectiveness of these tools is needed to ensure their effectiveness in the health care context. In conclusion, the transition to a territorial care system in Italy requires significant strengthening of municipal services, effective integration of health and social technologies, and a revision of funding policies to include innovative tools that improve patients' quality of life. Only through an integrated and targeted strategy will it be possible to adequately meet the growing demand for community health services and ensure continuous, quality care.

In **Norway** when a patient is transferred between levels of care several measures are put in place to ensure quality and continuity. The transfer can also involve several challenges. An emphasis is placed on interdisciplinary collaboration between hospital staff and municipal health services to ensure a comprehensive patient pathway. There is a need to strengthen holistic thinking and understanding of the challenges each level is faced with to improve collaboration. An example of recent interdisciplinary cooperation to improve prevention, treatment, and management of pressure sores is the case municipality nurses that work closely with hospital nurses and doctors thru video and digital assistance solutions to ensure a comprehensive approach to patient care. Patients may receive an individual plan or care plan a that outlines their needs for services and follow-up after discharge. A care plan is an essential tool in healthcare that helps to identify and document a patient's needs, establish goals for care, and plan interventions to meet those goals. The care plan is individually tailored to each patient and based on professional knowledge. A national scoring system (e.g. Individual-based Statistics for Nursing and Care Services - IPLOS) is used to determine the patient's needs. Patient and caregiver education/training is provided to prepare them for the transition from hospital to home. Training could be even more effective with the use of more digital solutions. A coordinator or a primary contact from the municipality is assigned to ensure that the patient receives the necessary services and follow-up after discharge. The use of electronic patient records contributes to the flow of information between different levels of care. There is room for improvement in transfer of records. The municipality and hospitals are not on the same platform, and information to be sent is chosen by healthcare professionals, sometimes old information is sent and sometimes things are omitted. This can in turn lead to misunderstandings and delays in care. The measures already in place help to reduce the risk of readmission, but a more comprehensive digital assistance solution package would ensure that more patients receive the follow-up they need in the transition from hospital to outpatient hospital services or to municipal health and care services.

In Eastern Europe, in **Romania**, ensuring the continuity in patients' follow-up is still a problem. Patients are discharged from hospital with recommendations regarding medical treatment, lifestyle changes and a management plan for follow-up, as well as an appointment for the first follow-up visit. They receive a discharge summary, which should be taken to their general practitioner, who is the only doctor who has the whole medical history of the patient. Based on the discharge summary, the general practitioner prescribes the medical treatment each month and at set intervals the patients are re-evaluated by a physician from an ambulatory department. The timing of follow-up visits depends on the recommendation made by the attending physician at discharge. Most of the time, it is difficult to respect the time intervals between follow-ups recommended at discharge due to the large number of patients or to the physical distance between home and hospital, these often make the therapy adjustments to be impossible at an early stage of the disease. The highest risk of readmission is among the elderly, especially those who live far from hospitals. They usually are admitted as emergency due to their clinical deterioration. Patients will also be supervised by the general practitioner, who can refer the patient to a specialist doctor sooner than originally planned in case an aggravation of the patient's pathology or possible complications are identified. Since there in no freely accessible staff available in Romania to help with the reintegration process after discharge, this task is often undertaken by the patient's family. Romania's remote patient monitoring market is less developed than that of many Western European countries. However, the nation is undergoing a digital transformation in its healthcare sector, driven by government initiatives and a growing awareness of the benefits of digital health solutions. Despite challenges such as infrastructure issues and lower technology adoption rates among the population, the demand for advanced remote patient monitoring systems is increasing giving hopes that soon doctors will be able to timely identify predictive factors of clinical deterioration to ensure treatment adjustments and to reduce the risk of readmission or visits to the hospital.

**Technology Integration**

The care pathway is augmented with IoT, AI and digital assistance technologies with a view of improving patient outcomes and quality of care. The hospital to home transition requires the integration of IoT devices for remote monitoring of daily life activities, vital signs, and medication of patients and AI models for identifying problems leading to rehospitalization and personalizing the post-discharge follow-up and intervention using digital assistants [20].

**Post discharge, remote patient monitoring** is needed to comprehensively and non-invasively track daily activities, including physical exercise, sleep patterns, movement habits, social interactions, and essential daily tasks like taking prescribed medications. This will help evaluate adherence to recommendations and medication plans. Figure 2 shows an example of such Remote Patient Monitoring platform and its associated services [21]. The platform provides a solid and secure basis for medical and care services running in the cloud. It includes a state-of-the-art identity broker for authentication, which can be configured to forward authentication requests to any other standards-compliant identity broker. Together with fully encrypted data storage and role-based access control, it fulfils all security requirements for the very strict e-health domain. The system has an extensive dashboard for healthcare and other professionals, with views tailored to the many different roles and services.

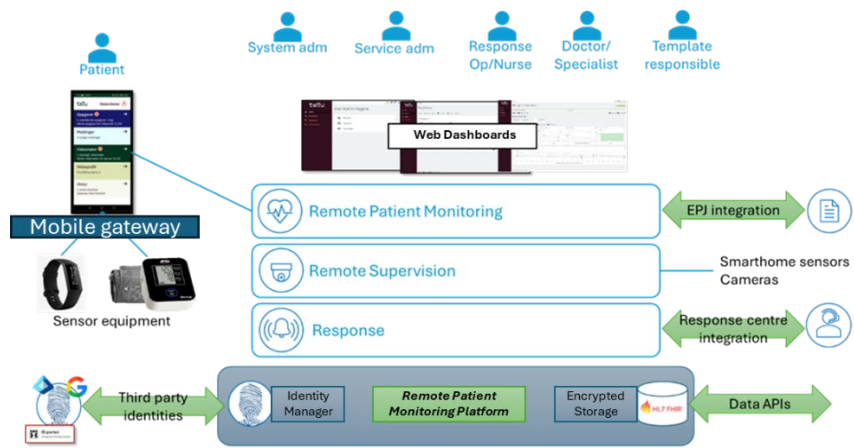

**Figure 2.** Remote patient monitoring infrastructure.

The patient uses a mobile app, that functions as a gateway for connecting sensor equipment and notifies the patient of their tasks. Off-the-shelf commercial wireless sensor devices are available to monitor the health, activity level and other relevant parameters but they provide data in heterogenous formats. Activity tracking devices such as modern fitness trackers and smartwatches seem to be a good solution for collecting various vital parameters and activity information, but do not typically allow direct integration. Additionally beacon tags and data from smart appliances can be used. Contextual information such as family and informal caregivers' perspectives and patient self-reporting can be collected with questionnaires and augment the sensor monitored data. The collected data can be viewed in the dashboards by users with the relevant permissions, and it can be made available to the ML through APIs.

**AI Analytics** are needed to process the collected patient monitored data, to establish a patient heath status baseline and identify events that indicate changes, whether sudden or gradual, that may lead to re-hospitalization. These changes may signal symptom progression, non-adherence to treatment, or a decline in well-being, potentially leading to re-hospitalization [22]. For generating insights over patients' states, AI technologies can take advantage of neural networks models to analyse the data monitored on daily life activity and vital signs to the identify problems that require proactive intervention and adherence to the prescribed treatment. AI pipelines (see Figure 3) deal with data pre-processing and cleaning, feature selection, cross validation, model training/evaluation and usage in care applications. They need to be customized and scaled to the transitional care specific context and to the data that is monitored in the system. A deep neural network technique utilizing Convolutional Neural Networks (CNN) models to analyse objective activity data from wearable devices such as fitness trackers, smartwatches, and smart patches seem to be a good solution for fitting remote care scenarios [23]. The valuable insights into the patient's state can be made available to physicians by integrating the ML results and predictions into the monitoring platform dashboards, leaving the decision to the healthcare personnel but aiding them in the complex decision-making process.

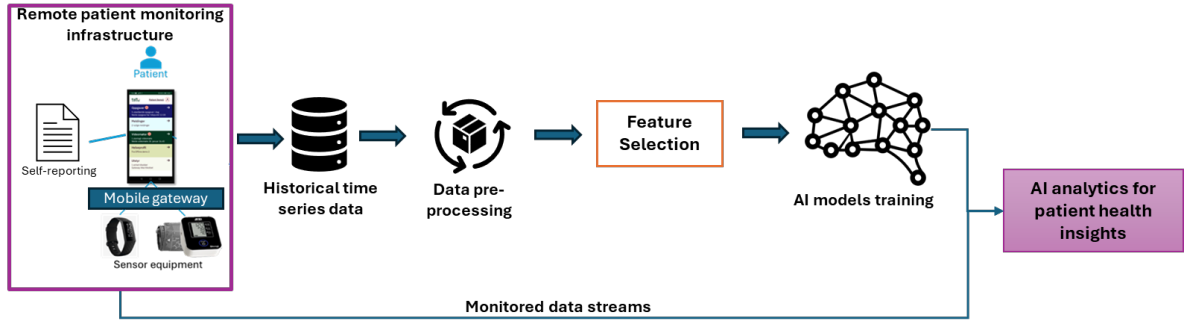

**Figure 3.** Pipeline for AI based patient healthcare status assessment

The coordination and follow up should enable smooth communication among stakeholders involved in the transitional care process. A dedicated **digital assistance communication platform** is needed to show patient health status information and will enable self-reporting of potential problems and wellbeing [24]. The platform may provide personalized nutrition advice for the patient based on the patient's diagnoses, training advice and information about the disease. It will also enable interaction and communication of all types of interested actors with the system, to offer support (reminders, step-by-step instructions etc.), and to allow configuration and personalization of the care process (see Figure 4).

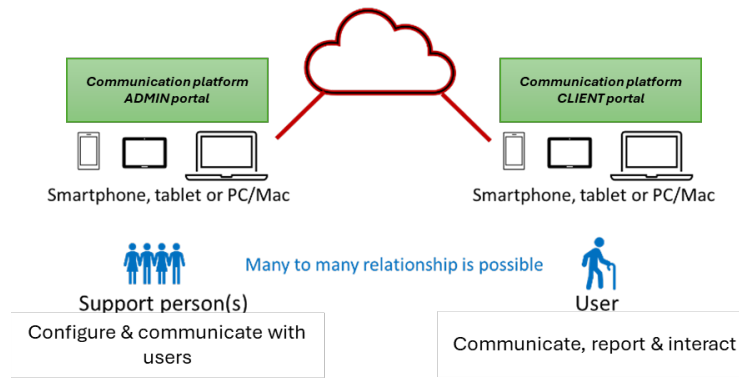

**Figure 4.** Communication platform overview.

**Design of validation trials**

The trial protocol for validating the envisioned enhanced care pathways needs to integrate primary, secondary, and tertiary users to cover the entire transitional care process. Additionally, these end-users can be also involved in the activities related to analysis of the transitional care process to identify the challenges facing health and care systems by involving different user profiles, the analysis of typical tasks or goals of patients, formal and informal carers as well as health professionals, and the deduction of principles and challenges for adopting the new proposed technologies. A close cooperation among healthcare stakeholders and technology providers will enable the adaptation of the transitional care technology, solving major usability and acceptability problems and addressing specific national healthcare settings contexts of specific pilots.

The trial main objective is to assess the potential of the technology to reduce the rate of rehospitalization within 30 days relieving the pressure on health and care facilities, aiming to establish a clinically validated value proposition for the technology. The technology will be deployed and tested in real condition of use, for a long-term to conduct a detailed impact assessment analysis aimed to understand patient and healthcare systems outcomes in terms of acceptability, quality of life, health improvement and cost reduction. During the field trials, support will be offered for installing the system and for training users to set up and use the transitional care services and monitoring infrastructure. A help desk needs to be available in each testing country and intervention needs to be guaranteed each time needed, ensuring a high motivation of patients, which influences the quality and the number of datasets strongly. Patients will be monitored through interviews on a fixed time basis. A technical review of the system design and functions to answer meaningful problems and difficulties in use which have arisen during the first months of the trial will be carried out. Finally, the field trial will be accompanied by a questionnaire which will be used to generate the dataset for the Cost-Benefit Analysis (CBA) [25]. CBA will focus on transitional care assistance after the introduction of the technology and compare this process across the welfare systems operating in the pilot countries. CBA will elaborate such institutional analysis by contrasting alternative models of welfare provision. In these settings it will examine whether technological change in the form of new devices can lead to savings for the welfare system at large as well as for individuals. A first concern will be the direct assessment of the level of satisfaction gained by using the developed technology and if this matches initial expectations. Subsequently, specific improvements in the quality of life and overall health due to the system, as well as information on hospital readmission after 30 days will be of interest.

To understand the effectiveness of the proposed enhanced care pathway in counteracting the readmission at the hospital after 30 days, a high number of patients needs to be enrolled over at least 3-months trial. The participants will be stratified based on potential predictors of 30-day readmission risk through the Comprehensive Geriatric Assessment (CGA) [26] before their discharge from the hospital. The study split participants into intervention and control groups. According to the study of Richie et. al. [27], 30-day rehospitalization rates between the intervention and control groups were 17.7% vs. 8.7% for chronic obstructive pulmonary disease and 22% vs. 12% for heart failure. Assuming these effect sizes, 80% statistical power, alpha error of 0.05 and using two-sided chi-square tests for equal proportions without adjustment for covariates, 221 subjects for each group should be recruited. In our study, involving older subjects with multimorbidity, we assume the same conditions and a potential drop-out rate of 20%.

Before the discharge, the healthcare professionals will assess the level or risk through the CGA and will suggest the use of the system to provide to the patients, through personalized contents and information aimed at supporting the self-management of their own health at home. Then, a preliminary assessment of attitude and eHealth skills, as well as goals to be fulfilled through the system will be conducted. Once installed, the participants will be remotely monitored and supported through the envisioned transitional care platform. After one month of use, a second interview will take place, to collect information on 30 days readmission, health and cognitive status, quality of life and usability of the system. After three months from the start of the trial, a final follow up will be conducted to detect the adherence rate of the participants to the socio-technical intervention and the overall satisfaction and usability.

## Discussion

The integration of IoT, AI and digital assistance technologies can have a significant long-term impact by enhancing the coordination and continuity of care, improving patient outcomes, and reducing the burden on healthcare systems [28]. However, a challenge to consider is the definition of a set of guidelines for re-designing the care paths for patients to consider the utilization of the novel technology developed in different care settings and national contexts [29]. The guidelines will consider the specific circumstances and healthcare systems of the countries where the trials are conducted, defining the roles and responsibilities of healthcare providers, caregivers, and patients to improve the coordination. This is important in Europe where the healthcare system is rather fragmented with specific rules and regulation in each country. The continuity of transitional care can be ensured [30] by leveraging on the innovative technology scaled, adapted, and validated in pilot trials ensuring remote patient monitoring post-discharge in their home, providing virtual links for doctors to follow up and gain insight to provide appropriate care levels based on the patient's medical condition with optimal allocation of resources will facilitate the sharing of data across the entire care continuum, which will include all interested stakeholders involved in the patient's care. This will foster better communication and collaboration among healthcare providers, ultimately enhancing the overall patient experience and satisfaction.

Additionally, the augmented care pathway will impact the healthcare systems by providing healthcare providers closer to real-time patient data, advanced ML-driven analytics to assess potential deviation from the set care plan, and enhanced communication across the transitional care continuum focused on hospital-to-home transitions [31]. The large volume of data generated through IoT devices and analysed using ML can uncover valuable insights, identify trends, and facilitate evidence-based personalized care [32]. Furthermore, the approach will lay the foundation for future EU care systems by redesigning the patient care paths in care transition from hospital to home considering the adoption of innovative technologies leading to more effective treatments, optimized resource utilization, and accelerated research and innovation in the field. Such developments can contribute to the important challenge of future standardization of transitional care models across healthcare settings by providing guidelines for care paths re-design promoting best practices that may facilitate successful transitions. However, open challenges still need to be addressed to ensure the seamless interoperability between healthcare systems, various actors in the system and the patient ensuring data privacy and security.

Due to its potential to improve the effectiveness of transitional care processes, the envisioned technological solution will have a strong direct social impact and can create significant value in the public sector. It can improve the patient's experience by ensuring a coordinated transition from hospital to home, leading to improved patient satisfaction with the healthcare system, better health outcomes, and the wellbeing of both the patient and the caregiver. From an economic perspective the innovation adoption into the existing healthcare systems can reduce the hospital readmission, thus contributing to the reduction of cost burden and better allocation of healthcare resources [33]. Simultaneously, it will help minimize the waste of healthcare resources by providing a comprehensive understanding of how to better anticipate and address issues faced by patients, family caregivers, and healthcare professionals during the transition from hospital to home. Another foreseen impact addresses the health management of high-risk populations, such as older adults with chronic diseases, offering them improved monitoring and communication support and improving the overall health outcomes. Thus, it will pave the way for the adoption of new patient centred and digitalized transitional care models while reducing the associated costs. However, the costs of implementation, and addressing potential resistance to adopting new technologies among healthcare providers and patients, need to be carefully managed.

Through the parameters they provide, remote patient monitoring devices have multiple potential clinical applications in cardiovascular pathology. Measuring physical activity levels can encourage lifestyle changes, having beneficial effects in both primary and secondary prevention of cardiovascular diseases. They allow the screening of arrythmias in high-risk patients, as well as the remote management of chronic diseases through the titration of drug treatment [34]. At-home monitoring of patients would allow reducing the length of hospitalization in the case of patients undergoing interventional procedures, such as patients with severe aortic stenosis treated percutaneously. These patients could benefit from an early discharge (in the first three post-procedural days or even the next day). Post-procedural arrhythmic complications, some of which can be life-threatening, could be detected with the help of wearable devices [35]. In a recently published study that evaluated the physical activity of patients who underwent percutaneously aortic valve replacement, with the help of smartwatches, it was observed that physical activity increases significantly in the first 2 months after the procedure and slowly decreases in those at risk of hospitalization. The increase in physical activity was also associated with a lower risk of 1-year mortality or rehospitalization especially in the case of those patients who took less than 5000 steps per day [36]. The technology-assisted transitional care models can help healthcare providers to share knowledge and communicate more effectively with patients. This can lead to lowering the rehospitalization percentages and more efficient allocation of healthcare resources. Finally, the new technologically enhanced care pathway will support several goals for sustainable development [37], helping to ensure healthy lives and promote well-being for all at all ages. It will strengthen the capacity of all countries, in particular developing countries, for early warning, risk reduction, and management of national and global health risks.

**Acknowledgement:** This work was supported by a grant (Nº: 1449, TransCare) co-funded by the European Union under the Transforming Health & Care Systems Programme and the Executive Agency for Higher Education, Research, Development and Innovation Funding (UEFISCDI) (RO), The Research Council of Norway (NO), and Italian Ministry of Health (IT). Views and opinions expressed are however those of the author(s) only and do not necessarily reflect those of the European Union, the National Agencies or European Health and Digital Executive Agency (HADEA). Neither the European Union nor the granting authorities can be held responsible for them.

**Author Contributions:** Conceptualization, I.A. and T.C.; Methodology, I.A. and T.C.; writing—original draft preparation, R.B., F.B., T.G., R.H., A.S., L.T.B., O.A., A.N. and C.G.; writing—review and editing, R.B., F.B., T.G., R.H., A.S., L.T.B., O.A., A.N. and C.G.; All authors read and agreed to the submitted version of the manuscript.

**Competing Interest:** The authors declare no conflicts of interest.

**Human Ethics and Consent to Participate declarations**: not applicable.

**Data availability:** All data generated or analysed during this study are included in this published article [and its supplementary information files].